\documentclass[11pt,a4paper]{article}

\usepackage[utf8]{inputenc}
\usepackage[T1]{fontenc}
\usepackage{amsmath,amssymb}
\usepackage{graphicx}
\usepackage[margin=1in]{geometry}
\usepackage{setspace}
\usepackage{hyperref}
\usepackage[super,comma,sort&compress]{natbib}
\usepackage{booktabs}
\usepackage{enumitem}
\usepackage{xcolor}
\usepackage{parskip}
\usepackage{wrapfig}
\usepackage{float}

\usepackage{booktabs}
\usepackage{amssymb}

\hypersetup{
colorlinks=true, linkcolor=blue!60!black, citecolor=blue!60!black, urlcolor=blue!60!black, pdfauthor={}, pdftitle={Emergent Fatigue Dynamics from Foundational Interatomic Potentials} }

\title{Emergent Fatigue Dynamics from Foundational Interatomic Potentials}

\usepackage{authblk}

\author[1]{Jingye Tan}
\author[1]{Benjamin A. Jasperson}
\author[2]{Eric Rothchild}
\author[2]{Anders Johansson}
\author[3]{R\`{e}mi Dingreville}
\author[1]{Krishna Garikipati}
\author[2]{Mitchell A. Wood\thanks{Corresponding author: mitwood@sandia.gov}}

\affil[1]{Department of Aerospace and Mechanical Engineering, University of Southern California, Los Angeles, CA, USA}
\affil[2]{Center for Computing Research, Sandia National Laboratories, Albuquerque, NM, USA}
\affil[3]{Center for Integrated Nanotechnologies, Sandia National Laboratories, Albuquerque, NM, USA}

\date{}

\begin{document}

\maketitle

\begin{abstract}
Foundational machine-learned interatomic potentials are trained on large, diverse DFT datasets, but their validation rests on equilibrium states, which does not guarantee correct emergent behavior in large-scale molecular dynamics.
Where no direct experimental validation is available, comparisons to physically motivated potentials are valuable for drawing insight.
We distinguish static properties, which are thermally accessible, from emergent properties, which require work from an external fatigue stress to drive rare events, and compare the potentials on both.
An EAM potential and the NequIP foundational potential both agree closely with reference data for many near-equilibrium properties, but NequIP predicts an anomalously low intrinsic stacking fault energy ($\sim7$ times lower than EAM) and thus predicts far more plastic deformation under fatigue cycling of a nanoscale flaw modeled after the classical Griffith crack problem.
Per-atom energy distributions after loading show that the two potentials differ most at energies far above what is thermally accessible.
We also document memory-related crashes and silent extrapolation as practical concerns for practitioners of these foundational model forms.

\end{abstract}


\section{Introduction}
Emergent dynamics of large ensembles of interacting particles are often difficult to predict from the equilibrium states of the system.
As a prime example, protein folding\cite{jumper2021highly,bertoline2023before} and drug design\cite{durrant2011molecular,friesner2004glide} have been notoriously challenging predictions to make due to the complex energetic landscape of each atom, chemical moiety, and solvent species.
Molecular dynamics (MD) simulations have proven to be valuable tools in this endeavor due to their ability to sample through states and conformations, including rare events and their local minima.
Advanced sampling methods such as metadynamics\cite{laio2008metadynamics,ceriotti2011simplifying} and umbrella sampling\cite{kastner2011umbrella} move beyond, toward enumerating the local minima that contribute to the partition function in order to predict the true observables of interest.
An analogous challenge exists for solid materials, where the strength of a plastically deforming metal can only be predicted with a large ensemble of atoms\cite{zepeda2017probing,zepeda2021atomistic}, and arguably an ensemble of grains in a representative microstructure\cite{wolf2022deformation,schiotz2003maximum,schiotz1998softening}.
Connecting properties measurable at equilibrium (thermally accessible states) to predictions of dynamic response of materials that includes high-energy local minima is a central challenge for computational material science.
For the community of MD practitioners, developing accurate and efficient interatomic potentials continues to be a grand challenge, and one to which advances in machine learning and AI have contributed greatly\cite{jacobs2025practical,eyert2023machine}.

Analytic forms of interatomic bonding were the first models used in MD and were only capable of studying noble gases\cite{lennard1937critical,paskin1972molecular,wang2020lennard}, simple covalent bonds\cite{finnis1989interatomic}, and fixed-charge ionic systems\cite{hugginsInteratomicDistancesCrystals1933}.
The central approximation shared by most classical potentials is that short-ranged interactions are all that dictate the energy and force on an atom.
The second generation of potentials can be classified as environment-dependent potentials and include such examples as EAM\cite{daw1993embedded}, bond-order\cite{brenner2000art, brenner2002second}, and more advanced extensions in MEAM\cite{jelinek2012modified}, variable-charge models\cite{rappe1991charge}, and ReaxFF\cite{van2001reaxff}.
Each of these model-form improvements incurs greater computational expense\cite{plimpton2012computational}, a trend that machine-learned\cite{zuo2020performance} and foundational interatomic potentials\cite{riebesell2025framework} have followed.
Rather than enabling larger or longer simulations, the gains from peta- and exa-scale supercomputing platforms have largely been consumed by this increase in computational cost, so typical atom counts and timescales have stagnated.
Exceptions to these trends exist; three machine-learned potentials (DeepMD, SNAP, Allegro) have been awarded or named as finalists for the ACM Gordon Bell Prize, which annually benchmarks scalability to leadership computing platforms\cite{jiaPushingLimitMolecular2020,nguyen2021billion,kozinsky2023scaling}.
The increase in compute cost given to (machine-learned) interatomic potentials (ML-IAP) is driven by the accuracy of said models, many of which have been proven to be quantum or chemically accurate, with energy errors approaching 1~meV/atom or 0.02~kcal/mol\cite{smith2019approaching,riebesell2025framework}.

While training errors for modern, purpose-built ML-IAPs are within the convergence of density functional theory (DFT) calculations, a central difficulty is validation against a ground truth for system sizes beyond the reach of electronic structure methods.
For (routine) MD simulations larger than a few thousand atoms, benchmarking directly against a full electronic structure calculation is not computationally feasible, as the cost of even conventional Kohn--Sham DFT scales cubically with system size\cite{mattsson2005designing}.
The central challenge is that properties of interest (defects, gradients, time evolution) are size-dependent results which obstructs the direct validation of MD to DFT.
Recent efforts have instead sought to bridge this gap by treating an ensemble of interatomic potentials as ``synthetic materials'' to learn correlations between DFT-accessible microscopic properties and large-scale quantities of interest; evaluating these microscopic properties with DFT then yields predictions of the large-scale quantities with quantified uncertainty\cite{jaspersonCrossscaleCovarianceMaterial2025, jaspersonFundamentalMicroscopicProperties2025}.
Direct comparison with experiment is also an imperfect solution.
Experimental observables are ensemble and time averages that can agree with simulation for the wrong mechanistic reasons, and typically obscure atomistic detail that motivates simulation in the first place\cite{dingreville2017primer}.
Parameter sweeps and trend extraction are comparable across methods but it is the mechanistic detail that is lost with imperfect comparisons.
Point-wise accuracy on energies and forces\cite{fu2022forces,sikorski2023machine} is also not a guarantee of correct emergent behavior, since errors in rarely visited regions of the potential energy surface will propagate to macroscopic predictions\cite{perez2024recent}.
New benchmarks are needed that probe emergent behavior of materials from the atomistic perspective.

Fatigue and fracture of metals offer such a benchmark.
They are emergent properties of the potential energy landscape\cite{barr2023autonomous,boyce2026materials}, driven by rare, high-energy events such as dislocation nucleation at a crack tip that are unlikely to be represented in training data, yet classical theories connect their outcomes to unit mechanisms that can be computed directly.
This work evaluates the equilibrium and emergent fatigue properties of copper predicted by a physically grounded EAM potential and an off-the-shelf NequIP foundational interatomic potential, as a benchmark for the viability of foundational IAP.
This effort will guide future fine-tuning exercises where rare events are deliberately sampled from MD to refine foundational models.

\section{Methods}
\label{sec:methods}
We target two interatomic potentials for comparison of the emergent dynamics that result in defect growth and ultimate mechanical failure of a cyclically loaded metal.
To make this comparison meaningful beyond final outcomes alone, we quantify three classes of properties: static properties from single-point calculations (e.g., elastic constants and stacking fault energies), thermally accessible properties sampled with MD (e.g., thermal expansion), and the non-equilibrium, path-dependent response to cyclic mechanical loading.
All calculations were performed using the LAMMPS molecular dynamics simulation engine; example input scripts and analysis tools will be hosted on the SciDAC MIRAGE project GitHub page \url{https://github.com/mirage-scidac}.
Further details on how to access and utilize these resources are given in the supplemental materials.

Our baseline interatomic potential is the Embedded Atom Method (EAM) from Mishin \textit{et al.}\cite{mishin2001structural}, which was chosen based on prior studies\cite{pei2015brittle,ogata2002ideal} that show good agreement with available experiments.
Notably, this interatomic potential has been used for low-temperature brittle fracture studies of nanotwinned materials\cite{zeng2015fracture} and ductile failure of polycrystalline structures\cite{cao2008grain}, which bracket the conditions studied here.
We compare this model against Neural Equivariant Interatomic Potentials (\texttt{NequIP}), a representative equivariant message-passing ML-IAP\cite{batzner20223}.
Here, species are first embedded into initial node features $h_i^{(0)}$.
Each of $T$ message-passing layers then (i) constructs messages on edges from the current features and the relative geometry, $m_{ij}^{(t)} = \mathcal{M}^{(t)}\!\bigl(h_i^{(t)}, h_j^{(t)}, \mathbf{r}_{ij}\bigr)$; (ii) aggregates them permutation-invariantly at each node, $m_i^{(t)} = \sum_{j\in\mathcal{N}_i} m_{ij}^{(t)}$; and (iii) updates the node features, $h_i^{(t+1)} = \mathcal{U}^{(t)}\!\bigl(h_i^{(t)}, m_i^{(t)}\bigr)$, with all operations constructed to preserve the correct rotational transformation behavior.
Per-atom energies are read out from the final features, and forces follow by differentiation with respect to atom positions.
Message passing has a computational consequence relevant to deployment: each layer propagates information one neighbor shell farther, so the effective receptive field exceeds the cutoff radius $r_c$, increasing both per-edge work and the communication required at domain boundaries in a parallel run.

We evaluate both the NequIP OAM-S and OAM-L models\cite{kavanagh2026fast} for computational cost and find that the OAM-S model is the only practical model form for studying fatigue response in Cu (given current resources).
The NequIP OAM-S model contains 0.6M free parameters and is trained on the OMAT24 training set\cite{barros2026open} before being fine-tuned on the sAlex and MPTrj\cite{schmidt2023machine,wang2023symmetry, dengCHGNetPretrainedUniversal2023} data sets.
In addition to the two-layer potential containing an input dimension of 192 (128 $l_0$ and 64 $l_1$ tensor features), the OAM-S model is built with core repulsion via the ZBL functional form\cite{ziegler1985stopping}.

\subsection{Computational Efficiency}
\label{sec:methods-eff}
Atomistic molecular dynamics simulations are linear scaling compute cost methods due to the short-ranged interactions of atoms, which map ideally to the domain decomposition parallelization schemes used in LAMMPS\cite{thompson2022lammps}.
Particles that cross the boundary of a domain are communicated to surrounding compute resources (MPI ranks) through ghost atoms that are appended to the compact neighbor lists handled by each rank.
The maximum interaction distance where forces are calculated is defined by the interatomic potential as the radial cutoff ($r_{cut}$) distance.
For the Cu EAM potential\cite{mishin2001structural} used here, $r_{cut}=5.507$\AA, and the graph features from the NequIP-OAM-S\cite{kavanagh2026fast} are constructed up to $4.50$\AA~away from the central atom.
As noted previously, message-passing neural networks aggregate atomic information beyond $r_{cut}$, which blurs this distinction with respect to empirical potentials.
In addition to these interaction distances, the computational complexity of the IAP functional form will determine the pre-factor to the linear scaling cost.
Evaluating the interatomic potential is the most expensive part of MD, and it is safe to assume that all other code segments in LAMMPS are equivalent between EAM and NequIP.

Two measures of computational efficiency are achieved by running short simulations of either potential type through LAMMPS and capturing the runtime.
First, a cubic cell of fixed size ($1.3\cdot10^{5}$ or $1.05\cdot10^{6}$~atoms; exact sizes in Table~\ref{tab:scaling-sizes}) is run for 1000 timesteps on progressively larger node counts, where each node contains 4 AMD MI250X GPUs (8 GCDs) (OLCF Frontier).
This is denoted as the strong scaling of the simulation and exposes the ideal workload (atoms/node) before communication between MPI ranks degrades performance.
By contrast, weak-scaling data correspond to a fixed ratio of atoms to nodes, with both scaled up in unison to ensure applicability to the problem sizes in the next section, where long times and large atom counts are needed.
The weak scaling is reported as the ratio of performance to the single node simulation of the same atoms/node; ideal weak scaling is then a value of unity.

\subsection{Equilibrium Property Measurement}
\label{sec:methods-eq}
Elastic constants are measured by applying small finite strains to the cell about its equilibrium state, obtained by minimizing the total energy with respect to atomic positions and cell dimensions, and computing the resulting change in stress.
With the equilibrium volume we also perform a series of volumetric expansions and contractions of the cell without modifying the fractional atom positions.
These curves define the zero temperature equation of state for the material and are a simple comparison between model forms.
Sampling and comparing these high symmetry states is a limited view of the predictive capability of either IAP.
A simple extension is then to address the thermally accessible states of FCC Cu.
We do so by running an isobaric (P = 1~atm) simulation for 20~ps over the temperature interval of [0,1000]K in 100K increments.
The average cell volume is recorded at each temperature, using only the last 10~ps to make the measurement.

Beyond these high-symmetry states of FCC metals, the lowest energy pathway for these materials to alleviate an externally applied stress is to nucleate stacking faults and dislocations.
The energy barriers for the formation and motion of these one (dislocations) and two (stacking faults) dimensional defects are calculable in a small volume of material by displacing planes of atoms along specified high-symmetry directions.
A semi-periodic slab of material is oriented such that the non-periodic (z-) direction is normal to the stacking fault plane and one of the orthogonal (x-, y-) directions contains the close-packed direction along which one half of the cell is displaced.
For an FCC metal the z-direction is [111] and displacements are made along the $[1\bar{1}0]$ and $[11\bar{2}]$ directions to fully reconstruct the stacking fault energy.
The barrier from the perfect lattice to a displacement of $\frac{b}{6}\langle 112\rangle$ is known as the unstable stacking fault energy (USFE) and the energy difference between the perfect lattice and a displacement of $\frac{b}{3}\langle 112\rangle$ is the intrinsic stacking fault energy (ISFE), where $b = 3a_0/\sqrt{6}$ is the repeat distance along $\langle 112\rangle$ in the (111) plane (Fig.~\ref{fig:gsf}); the ISFE displacement, $b/3 = a_0/\sqrt{6}$, is the Shockley partial Burgers vector $\frac{a_0}{6}\langle 112\rangle$.
USFE and ISFE are well-defined states on the potential energy surface of either model, but equilibrium dynamics rarely visits them.
Both are energies per unit area, so the cost of a fault grows with the area that slips.
Per atom in the (111) plane, the USFE is only $\sim$0.06~eV, roughly $2k_BT$ at 300~K.
A stacking fault, however, requires every atom on the slipped patch to move together, so its barrier scales with the number of atoms involved.
Thermal fluctuations alone therefore cannot create a fault.
An applied shear stress must do most of the work, and the USFE sets the scale of that stress.
Energy minima at each displacement are achieved through constrained dynamics of the atoms where only z-direction motion is permitted and in-plane (111) motion is only permitted through the external constraint.
Displacements are done serially where the previous state serves as the initial condition, saving on compute time and avoiding spurious local minima that hinder the conjugate gradient minimization solver used in LAMMPS.
Displacements are carried out over one full period in each direction to ensure the path is symmetric.
Total energy and atomic forces are both converged to a relative tolerance of $10^{-12}$, which is achieved within 100 conjugate gradient steps.

\subsection{Strain Controlled Fatigue}
\label{sec:methods-fatigue}
Our main comparison of the emergent dynamics of these two interatomic potentials comes from the mechanical response of bulk Cu containing a single idealized flaw.
The simulation is modeled after the well-studied Griffith crack geometry \cite{griffithVIPhenomenaRupture1921}, which is used to illustrate the balance between bulk cohesive energy, stored elastic energy, and surface energies within a crystal.
For this study, three principal tensile axes are designated, [100], [110], and [111], which means that our simulated crystal will be rotated with respect to the Cartesian simulation boundaries.
To begin, a unit cell is replicated to achieve a 50~nm $\times$ 50~nm $\times$ 4~nm volume that is periodic in all directions.
In the isobaric ensemble, the system is equilibrated to a set pressure in each of the Cartesian directions ($\sigma_{x,y,z}$) and a temperature through a Berendsen thermo-/baro-stat over 50ps.
With the tensile axis set as x, the crystallographic directions mapped to the Cartesian cell directions are i) x=[100], y=[010], z=[001]; ii) x=[110], y=$[1\bar{1}1]$, z=$[1\bar{1}\bar{2}]$; and iii) x=[111], y=$[1\bar{1}0]$, z=$[11\bar{2}]$.
These tensile axes resolve the applied load onto the $\{111\}\langle110\rangle$ slip systems with differing Schmid factors.
Shear stress components were not controlled, so the cell remained orthorhombic throughout; for orientations i) and iii) this yields a uniaxial stress state.
For orientation ii), elastic anisotropy and lateral cell faces that are not crystallographic mirror planes produce an additional shear stress that supplements the resolved shear stress on the active slip systems.
During this equilibration, the strained system stores elastic energy, which is then relieved by introducing a planar flaw.
To create this flaw the neighbor lists of adjacent planes of atoms are adjusted such that material above and below no longer interact with one another, thereby severing any bonds.
The initial width of the crack is 10\% of the y-length, regardless of the crack orientation, and the crack extends through the entire z-direction.
The system is re-established at the desired stresses over 200ps before neighbor lists are restored.
At this point, a small amount of plastic deformation is expected, which stabilizes the flaw against reforming as a bulk solid.

Continuation of these simulations from the initialized Griffith's crack geometry can proceed in two fashions, based on what control variables are of interest.
Deformation of the cell can be carried out to either a defined stress or strain amplitude from the starting configuration.
While experimental fatigue studies\cite{barr2023autonomous} operate in either mode, it is commonplace in the simulation literature\cite{wilson2019continuum} to use strain-controlled deformation; we adopt this as well.
A 0.5\% strain amplitude is applied with an oscillating cell length of period 1ns while lateral dimensions are held at a fixed pressure (see Figure~\ref{fig:render}A). \
Temperature is maintained through a Nos\'{e}--Hoover thermostat throughout the deformation cycle.
This cyclic deformation is repeated ten times, and between cycles the system is restored to the target pressures over 100ps.
Re-equilibrating the system ensures the target pressure remains at the median stress on the cell during subsequent cycles.
A 1~fs integration timestep is used throughout this protocol, with global state variables (stress, temperature, etc.) stored every 1~ps and local atomic information (strain, potential energy) every 20~ps.
Visualizations of the simulation cells are carried out in OVITO\cite{stukowski2010visualization}.

Two classical results organize the emergent observables of this study.
The Griffith energy balance states that a crack advances by cleavage when the elastic energy released per unit area of advance exceeds the cost of the two surfaces created, $2\gamma_s$; it defines the brittle limit and ties crack advance to the surface energy.
At an atomistically sharp tip, cleavage competes with the emission of a partial dislocation on an inclined slip plane, whose barrier is set by the unstable stacking fault energy ($USFE$).
In the Rice framework this competition is governed, up to orientation-dependent geometric factors, by the ratio $\gamma_s/USFE$: large ratios favor emission and blunting (ductile response), small ratios favor cleavage (brittle response)\cite{rice1992dislocation}.
Temperature enters through thermal activation of the emission event, so sweeping temperature in MD enables a comparison of the thermally accessible mechanisms of each potential.
The practical significance for this study is that $\gamma_s$ and $USFE$ are \emph{unit properties}; the Rice ratio therefore converts each potential's unit-scale energetics into an \textit{a priori}, falsifiable prediction of its large-scale mechanism map.
Sampling loading conditions that apply differing Schmid factors on preferred slip planes also amplifies the differences between potentials, as non-ideal slip will involve high-energy transition states.
As such, we define a set of boundary conditions to run strain-controlled fatigue deformation simulations at T = (50, 100, 200, 300, 400, 500)~K, tensile axis = ([100], [110], [111]), $\sigma_x$=[2,3,4,5]GPa, $\sigma_{y,z}$=$1\cdot10^{-4}$GPa.
Positive stress in this study indicates a tensile loading direction; note this is opposite of the LAMMPS output convention.

\section{Results \& Discussion}
\begin{figure}[t]
  \centering
  \includegraphics[width=0.5\linewidth]{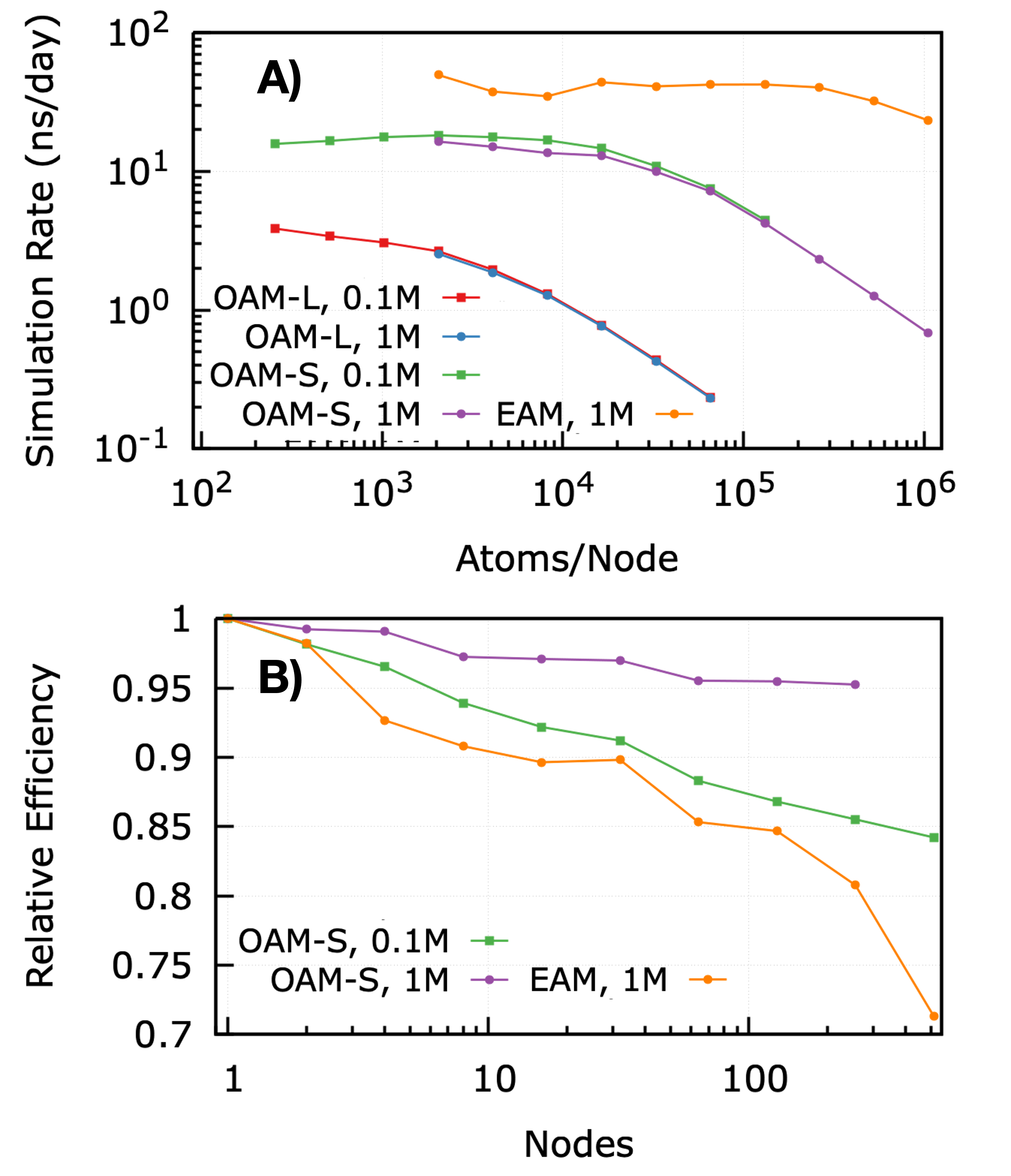}
  \caption{LAMMPS MD simulation performance for each of the potentials tested; NequIP-OAM-L was also run for additional comparison. A) Strong scaling data measuring simulated time per wall-clock day for fixed problem sizes. B) Weak scaling, where the simulation size is scaled up as node count increases. Thus, the 0.13M data are per node and the largest system contains 67M atoms on 512 nodes. Legend labels 0.1M and 1M denote 131,072 and 1,048,576 atoms, respectively (Table~\ref{tab:scaling-sizes}).
  \label{fig:perf}}
\end{figure}

The results presented here compare the two potentials across equilibrium, thermally sampled, and fatigue-driven states to identify where NequIP interpolates and where it extrapolates on the potential energy surface (PES).
In addition to the accuracy of the underlying predictions, the computational cost of an empirical potential versus a foundational one affects what can be achieved.
Figure \ref{fig:perf} captures two measures of computational efficiency of each potential: strong and weak scaling on modern hardware.
Strong scaling data in Fig.~\ref{fig:perf}A) is gathered on a fixed-size problem (1.05M or 0.13M atoms) and run on 1 to 512 nodes of Frontier.
While the NequIP-OAM-S model is used for all comparisons to EAM, timing data relative to NequIP-OAM-L is presented as well.
A plateau in simulation rate, reported as simulated time over wall clock time, is non-ideal to plan simulations around, as these jobs are throttled by communication among nodes rather than evaluation of the IAP.
Even at 1M atoms per node, EAM does not show a linear tradeoff between simulation rate and resource usage, indicating that there is no advantage to utilizing leadership computing resources for the problem sizes studied here.
For both problem sizes (0.13M and 1.05M atoms), NequIP models are slower than EAM and show ideal workloads around $10^4$--$10^5$ atoms/node when trying to conserve overall node$\cdot$hour usage.
Additional data of communication versus IAP evaluation is available in the Supplemental Information.
From these data, it can be seen that for a properly load-balanced (linear trade-off of performance to resources) simulation, NequIP-OAM-S is 5.7 times slower than EAM, and OAM-L is an additional 31 times slower in this measure of strong scaling.

Weak scaling results in Figure \ref{fig:perf}B) show that the communication overhead of using more nodes affects EAM more strongly than NequIP models.
However, OAM-L was unable to allocate enough memory to load the model and its parameters at a fixed ratio of 131k atoms/node, which is also seen in the strong scaling data, where all reported data are below this ratio.
Because of this inability to reduce resource usage, OAM-S is the only NequIP variant used for the remainder of this work.
We also noticed performance degradation using NequIP-OAM-S when nodes ran out of memory unexpectedly.
This occurs because each node has to store the NN and its parameters, which is a static memory requirement, but the workload for each MPI rank will be altered by the deforming material.
Dynamic workload balancing available in LAMMPS was utilized in this study and tempered some but not all of these crashes.
Jobs that exited early required multiple restarts to achieve multiple fatigue cycles.

Using the performance data in Fig.~\ref{fig:perf} and fatigue simulation details given in Section \ref{sec:methods-fatigue}, we calculate the expense of a single NequIP fatigue simulation to be $\sim10^3$ node$\cdot$hours on the Frontier hardware.
Multiplying through the sampling grid of free parameters results in a total cost of $1.1\cdot10^3$ node$\cdot$hours for EAM fatigue simulations and $7.2\cdot10^4$ node$\cdot$hours for NequIP.
The EAM runs do not require leadership computing and were run on a local GPU cluster containing four NVIDIA A100s per node.
These results highlight that while computing resources are abundant, care must be taken when utilizing these foundational interatomic potentials, as their recent arrival has not left time for code optimization in LAMMPS.
Recent efforts are closing this computational efficiency gap\cite{kavanagh2026fast, li2026universal}, enabling wider application of these model forms.

\subsection{Equilibrium Properties}

The distinction between near-equilibrium and non-equilibrium sampling can be made quantitative through the Boltzmann distribution.
At temperature $T$, an equilibrium system visits a configuration of potential energy $E$ with probability proportional to
\begin{equation}
  p(E) \;\propto\; g(E)\, e^{-E/k_B T},
  \label{eq:boltzmann}
\end{equation}
where $g(E)$ is the density of states and $k_B$ is the Boltzmann constant.
States lying more than a few $k_B T$ above the local minimum are exponentially suppressed: at $T = 300\,\mathrm{K}$, $k_B T \approx 25.9\,\mathrm{meV}$, so a configuration $1\,\mathrm{eV}$ above the ground state carries a Boltzmann factor of order $e^{-40} \sim 10^{-17}$ and is, for practical purposes, never visited in equilibrium.
In an equilibrium constant-temperature simulation, the instantaneous potential energy fluctuates about its mean $\langle E \rangle$ with a variance fixed by the heat capacity,
\begin{equation}
  \sigma_E^2 \;=\; \langle E^2 \rangle - \langle E \rangle^2 \;=\; k_B T^2 C_V,
  \label{eq:fluctuation}
\end{equation}
so a plain isothermal trajectory of the pristine crystal directly measures the width of the thermally accessible states for each potential\cite{frenkel1997understanding}.
Additionally, the distribution of per-atom energies $E_a$ in an equilibrium crystal is narrow and unimodal, with a width set by $k_B T$; its mean and width define a reference band against which any defected or loaded configuration can be compared.
An atom whose energy lies many $k_B T$ above this band occupies a state that equilibrium sampling at that temperature would essentially never produce, and at which the two potentials are least constrained to agree.
While small probabilities do not constitute absolute outcomes, one must also consider the limited sampling time in MD for observing such a rare event.

Regarding states near equilibrium, Table \ref{tab:unit-props} collects basic material properties of both potentials against available literature values from DFT and experiments.
We see general agreement of both model types with these comparative data, though there is some disagreement regarding the predicted surface energies and ISFE.
\begin{table}[t]
  \centering
  \caption{Near-equilibrium unit properties of Cu for both potentials against reference values. Note referenced data does not conform to uniform methodologies for either experimental or DFT studies.
  }
  \label{tab:unit-props}
  \small
  \begin{tabular}{lcccc}
    \toprule
     & EAM  & NequIP-OAM-s & DFT & Experiment\\
    \midrule
    $a_0$ (\AA) & 3.615 & 3.624 & 3.62\cite{mishin2001structural} & 3.615\cite{mishin2001structural} \\
    $C_{11}$ (GPa) & 169.878 & 177.361 & 189.7\cite{Jona2001} & 170.0\cite{mishin2001structural} \\
    $C_{12}$ (GPa) & 122.586 & 142.141 & 128.2\cite{Jona2001} & 122.5\cite{mishin2001structural} \\
    $C_{44}$ (GPa) & 76.206 & 81.013 & 94.4\cite{Jona2001} & 75.8\cite{mishin2001structural} \\
    $\alpha_L$ ($10^{-6}$/K, 300 K) & 17.53 & 20.77 & 17.68\cite{Pike2019} & 16.61\cite{White1973} \\
    $USFE$ (mJ/m$^2$) & 162.910 & 165.814 & 158\cite{Zhang2019} &  \\
    $ISFE$ (mJ/m$^2$) & 44.553 & 6.382 & 39\cite{Zhang2019} & 45\cite{Zhang2019} \\
    $\gamma_{s}\,\{100\}$ (mJ/m$^2$) & 1345.3 & 1433.9 & 1783\cite{Han2019} & 1790\cite{mishin2001structural} \\
    $\gamma_{s}\,\{110\}$ (mJ/m$^2$) & 1475.5 & 1551.4 & 1892\cite{Han2019} & 1790\cite{mishin2001structural} \\
    $\gamma_{s}\,\{111\}$ (mJ/m$^2$) & 1239.5 & 1281.3 & 1292\cite{Han2019} & 1790\cite{mishin2001structural} \\
    $E_{coh}$ (eV) & -3.540 & -3.443 & -3.54\cite{mishin2001structural} & -3.54\cite{mishin2001structural} \\
    $E^f_{vac}$ (eV) & 1.276 & 1.064 & 1.272\cite{Sivak2007} & 1.27\cite{mishin2001structural} \\
    $E^f_{oct}$ (eV) & 3.281 & 3.341 & 3.243\cite{Sivak2007} & 2.8-4.2\cite{mishin2001structural} \\
    $E^f_{tet}$ (eV) & 3.298 & 3.214 & 3.553\cite{Sivak2007} & \\
    \bottomrule
  \end{tabular}
\end{table}
Cohesive ($E_{coh}$) and point defect formation energies ($E^f_{...}$) are included for completeness of static material property comparisons.
These are not single point measurements as they require the comparison of defect and pristine states.
Notable differences between EAM and NequIP cohesive and vacancy formation ($E^f_{vac}$) energies are in contrast to the close agreement in the octahedral ($E^f_{oct}$) and tetrahedral ($E^f_{tet}$) interstitial formation energies.
While these formation energies are far from thermally accessible ($k_B T \simeq 25$meV) at room temperature, they are high-symmetry defects that should be reasonably contained in the NequIP training set.
\begin{figure}[t]
  \centering
  \includegraphics[width=0.5\linewidth]{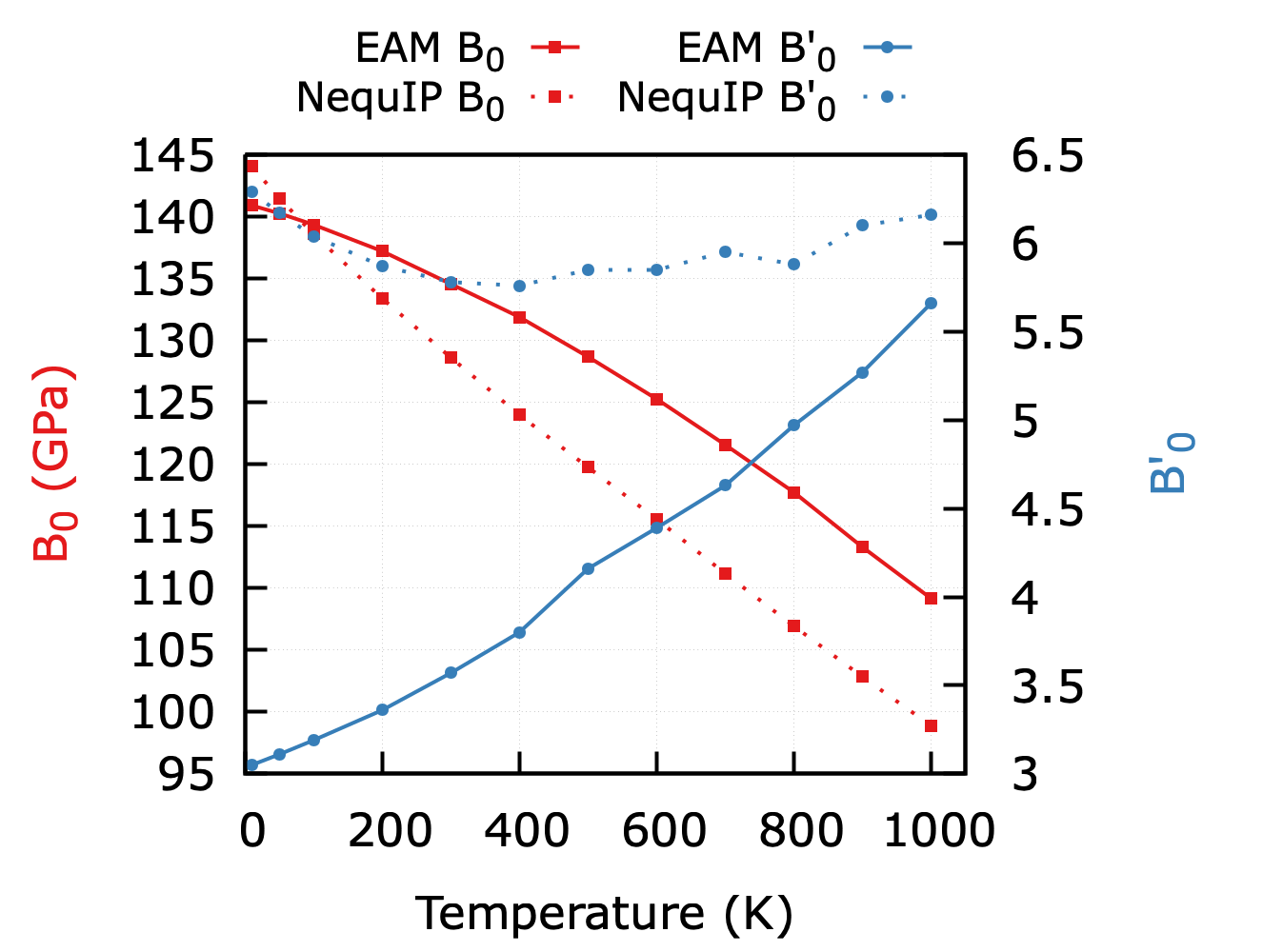}
  \caption{Bulk modulus and its pressure derivative from Birch-Murnaghan equation of state fits to isobaric (1~atm) simulations of either potential form.}
  \label{fig:expand}
\end{figure}

The E(V,T) data from the linear thermal expansion ($\alpha_L$) calculations enable a fit to the Birch-Murnaghan equation of state, which yields the bulk modulus ($B(T)$) and its pressure derivative ($B'(T)$), displayed in Figure~\ref{fig:expand}.
We carry this out up to $0.75T_m$ (EAM, $T_m=1326$K\cite{broughton1986thermodynamic}).
These are finer scale comparisons between the equilibrium states of either potential and yet still show meaningful differences.
With increasing temperature, the two potentials show similar decreases in the bulk modulus ($B_0$).%
Stark differences arise in the pressure derivative of $B_0$.
The low-temperature $B'_0$ from EAM ($\approx3$) lies below the value of $\approx4$ expected for most metals and below the experimental value for Cu at room temperature ($4.86$)\cite{gerward1985bulk}, and it rises steadily with temperature.
NequIP predicts a nearly constant $B'_0\approx6$, somewhat above the experimental value.

\begin{figure}[t]
  \centering
  \includegraphics[width=1.00\linewidth]{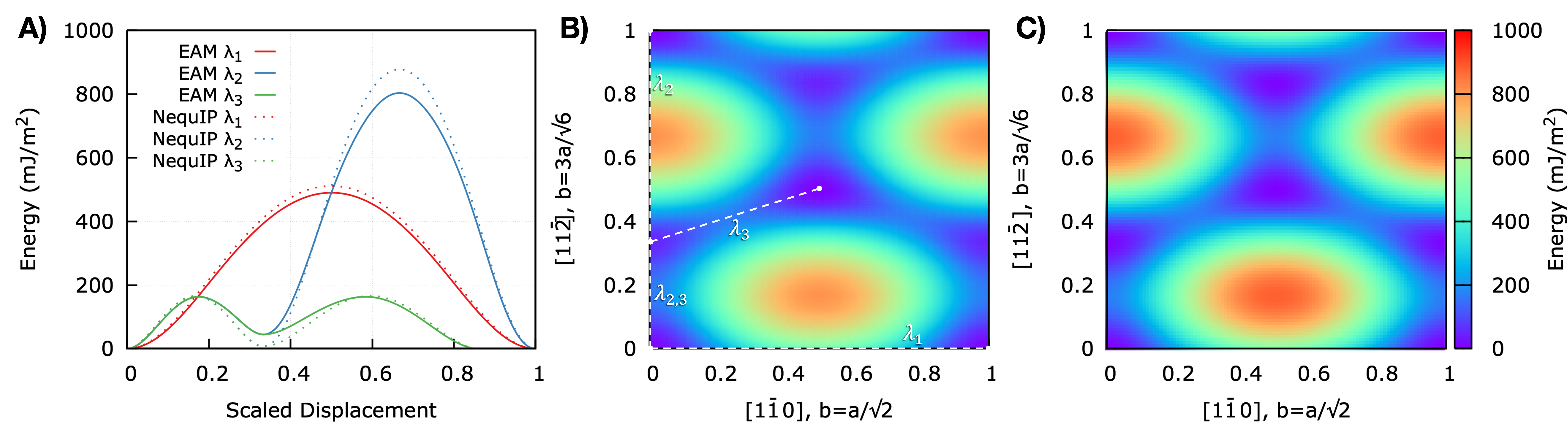}
  \caption{Full stacking fault plane energies and isolated low energy deformation pathways for B) EAM and C) NequIP-OAM-S. A) Full edge dislocation pathways $\lambda_{1,2}$ follow axis edges while partial dislocation dissociation $\lambda_3$ highlights changes in USFE and ISFE between potentials.}
  \label{fig:gsf}
\end{figure}

In FCC metals, dislocation-mediated plasticity dominates, as it is energetically favorable over twinning and surface (void) creation.
Rather than full dislocations, partials of $\frac{a_0}{6}\langle 112 \rangle$ character that sum to a complete $\frac{a_0}{2}\langle 110 \rangle$ Burgers vector are the energetically favorable pathway for FCC materials to deform, and capturing them is important for an interatomic potential.
To capture the partial dislocation nucleation barrier we perform a full stacking fault energy calculation by enumerating displacements in the (111) plane along the principal directions $[11\bar{2}]$ and $[1\bar{1}0]$.
Figure \ref{fig:gsf} displays the full stacking fault energies, panels B) and C), as well as the low energy deformation pathways in panel A).
Full edge dislocation pathways are shown as $\lambda_{1,2}$ in all panels, which do not pass through the low-energy paths in the stacking fault plane and have energy barriers several hundred mJ/m$^2$ higher than the USFE.
The partial dislocation dissociation pathway of a dislocation with an $\langle 110\rangle$ Burgers vector is shown as the low-energy pathway extending from the ISFE to the high symmetry point (0.5$[1\bar{1}0]$, 0.5$[11\bar{2}]$) that is equivalent to a stacking fault.
Of particular note between the EAM and NequIP GSF data is the similarity of the USFE (first peak along $\lambda_{2,3}$), which differs by only 1.8\%, while the ISFE (first minimum along $\lambda_{2,3}$) differs by a factor of seven.
The energy barrier for partial dislocation nucleation is given by the USFE, though much of this energy is recovered at the ISFE minimum; $USFE - ISFE = 118.357$~mJ/m$^2$ for EAM and $159.432$~mJ/m$^2$ for NequIP.
Wu \textit{et al.} discuss and compares the ductility of FCC metals through the ratio of USFE/ISFE, where higher ratios trend toward partial dislocations as the preferred deformation mechanism.
The abnormally low ISFE predicted by NequIP yields a ratio over three times higher than that of Ag ($USFE/ISFE = 6.53$), which was the highest value reported\cite{wu2010ab}.
Our EAM potential agrees well with the DFT values of Wu \textit{et al.} on Cu where $USFE/ISFE = 3.7$ and $4.1$, respectively.
Neither potential predicts cleavage along crystal planes as energetically more favorable than dislocation nucleation; see Table \ref{tab:unit-props} for the surface energy values ($\gamma$).
The Rice ratio $\gamma_s/USFE$ is nearly identical for the two potentials (7.6 for EAM and 7.7 for NequIP on \{111\}), so both are predicted to emit dislocations rather than cleave, as observed.
The differences in fatigue outcome therefore arise beyond the nucleation step, where the sevenfold difference in ISFE governs how faults evolve.
This observation of overly ductile behavior of NequIP OAM-S will shape much of the interpretation of the fatigue simulations.

\subsection{Fatigue Outcomes}
\begin{figure}[t]
  \centering
  \includegraphics[width=1.00\linewidth]{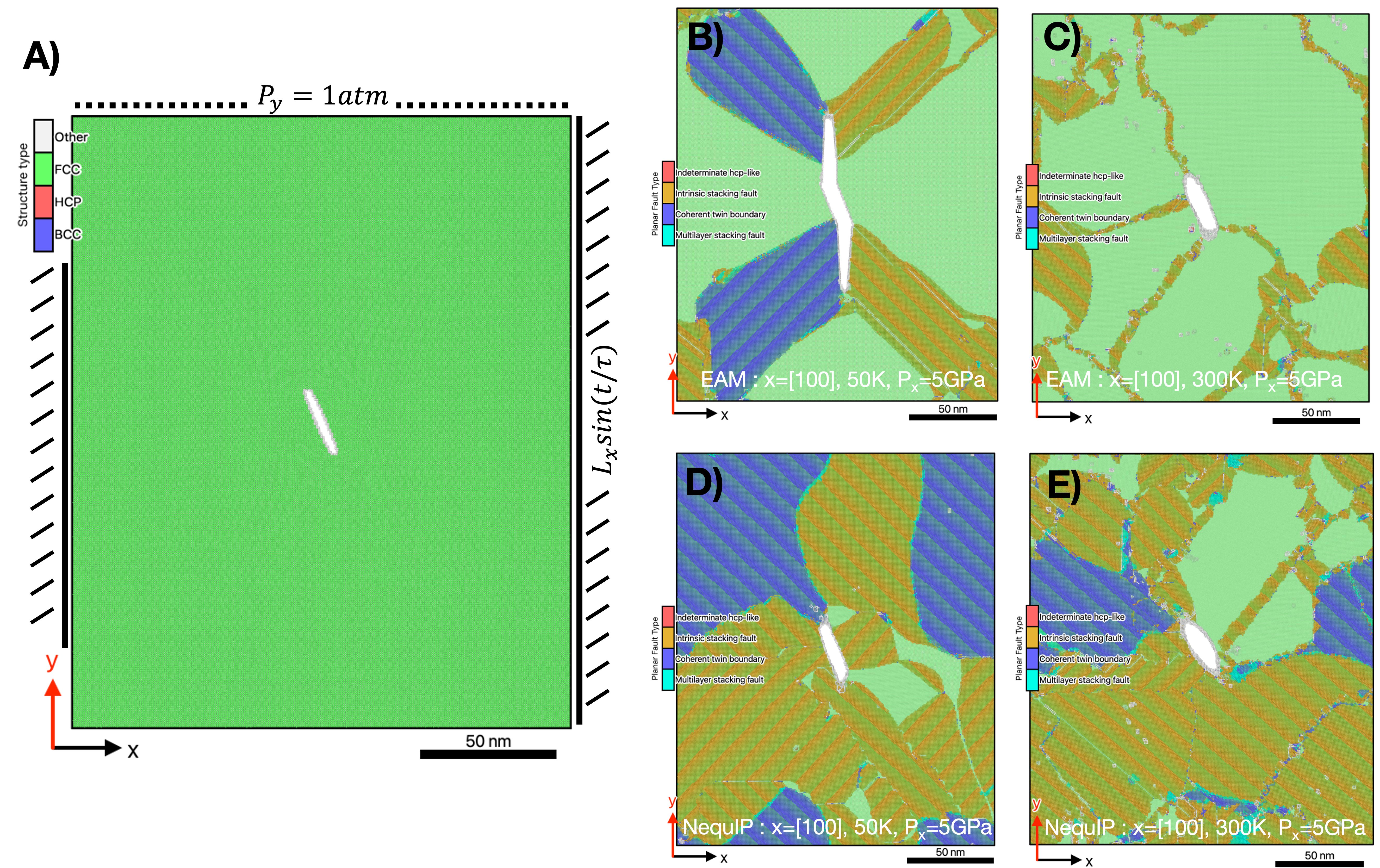}
  \caption{A) Initial condition of Griffith's crack simulations identifying oscillatory tensile direction and isobaric boundaries. EAM predictions of B) low-temperature and C) room-temperature deformation, where the tensile axis is [100] and the pre-stress value is 5~GPa. D) and E) match these conditions but utilize the NequIP OAM-S foundational potential.}
  \label{fig:render}
\end{figure}

Our primary means of comparing the two interatomic potentials is through the cyclic deformation of a bulk material with a single idealized flaw.
Figure \ref{fig:render} A) shows the initial condition of this set of simulations.
The initial FCC crystal has a crack oriented at 45$^\circ$ to the tensile axis, as identified by the white `other'-typed atoms in the graphic.
As deformation proceeds, the crack geometry evolves in a manner unique to the loading and thermodynamic conditions.
Panels B)--E) of Figure~\ref{fig:render} add a transparent effect to FCC atoms to highlight the planar fault types that originate from the crack surface.
The thin z-dimension wraps stacking faults and full dislocations into the observed `stripes'.
Low-temperature panels B) and D) show both potentials predicting crack propagation primarily along the y-direction mediated by partial dislocation emission from crack tips.
By contrast, room-temperature runs (panels C) and E)) show the flaw geometry being blunted as deformation progresses, though small differences in the flaw geometry can be observed at the minimum/maximum in the y-direction.
Figure \ref{fig:render} D) also shows some blunting of the crack tip though the geometry of the flaw as measured by the surface area to volume ratio highlights subtle differences.
These single snapshots in time do not reveal much about the fatigue outcomes predicted by either potential other than to demonstrate differences from classical crack propagation modes.

\begin{figure}[t]
  \centering
  \includegraphics[width=0.50\linewidth]{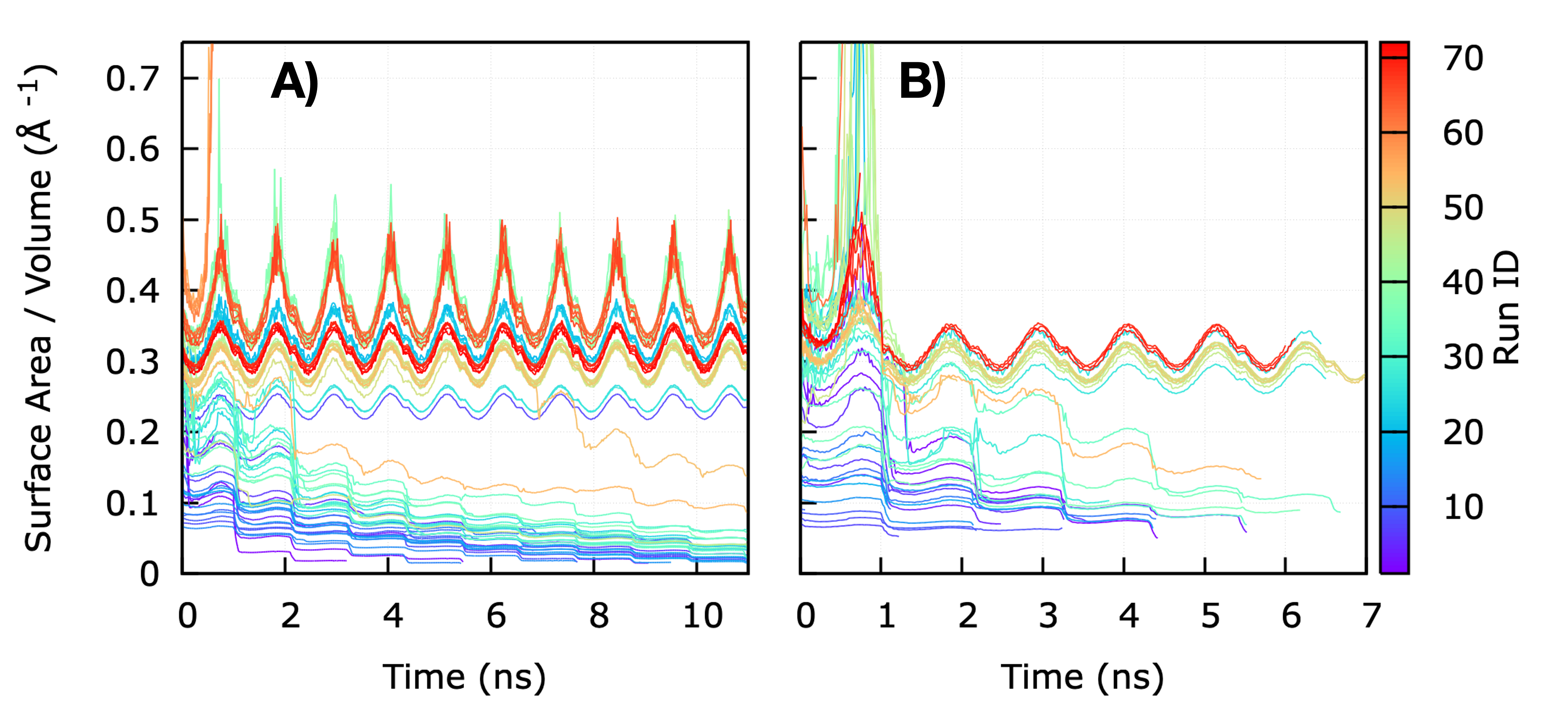}
  \caption{Evolution of the crack geometry for each of the sampled conditions utilizing A) EAM and B) NequIP-OAM-S. Run IDs are tabulated in the supplemental material sweeping through tensile orientation, temperature, and pre-stress values. }
  \label{fig:sav}
\end{figure}

A total of 72 fatigue simulations were conducted for each of the interatomic potential types spanning tensile orientation, temperature and pre-stress values.
From this batch of simulations we are able to discern where these two potential types differ beyond the equilibrium property predictions discussed earlier.
The first of these quantitative metrics is the geometry of the evolving flaw under dynamic load; here, we take the ratio of surface area to volume (SA/V) to indicate where dislocation emission blunts the crack tips.
Both panels of Figure \ref{fig:sav} use a color scale to indicate each unique simulation, the full details of which are available in Supplemental Table 1.
There are three distinct outcomes of the fatigue loading that can be identified in these figures.
First, where SA/V evolves toward zero, the flaw grows into a through crack, which is observed in 32 EAM and 28 NequIP simulations.
In addition to the similar number of fracture outcomes (denoted as `F' in Supplemental Table 1), the parameter bounds where this is observed are similar in both model forms.
These cases are isolated to high pre-stress values ($\geq4$~GPa), or to high temperatures ($\geq400$~K) at low pre-stress values.
Most simulations that observed this outcome did so within the first few cycles ($\leq3ns$) though rare cases (EAM IDs 53 and 54) show a late transition toward through cracks, indicating a threshold for flaw progression at these conditions.
A large number of simulations preserve the surface area to volume ratio throughout the fatigue cycles, only showing variation due to volumetric deformation of the cell.
These account for 34 EAM and 13 NequIP simulations.
We explain this large discrepancy in nominal (labeled `N' in Supplemental Table 1) outcome counts through the equilibrium property measurements in Table \ref{tab:unit-props}.
The NequIP model of Cu predicts an abnormally low ISFE, which means that dissipation of mechanical work will result in a proportionally higher rate of partial dislocation emission at the surface of the flaw.
Where the work done by the resolved shear stress during the tensile half-period exceeds the USFE per unit area of slipped plane, multiple nucleation events are expected (see Figure~\ref{fig:render}), but an excess of elastic energy remains due to the low ISFE relative to USFE.
Lastly, there is a third scenario where the flaw heals during the deformation cycle due to the pre-stress value being too low to sustain the geometry during the compressive half-period.
These are distinguished by SA/V diverging toward $+\infty$ before the flaw is erased and both quantities become zero; to suppress noise, we omit data for flaw volumes below 1~nm$^3$.
This last scenario accounts for only 6 EAM and 31 NequIP simulations.
Comparing where healing outcomes were observed between the two models, EAM only observed this where $\sigma_x=2$~GPa and the tensile axis was aligned with [100], with no temperature dependence.
By contrast, NequIP showed flaw erasure in each of the tensile directions and up to $\sigma_x=4$~GPa.
While we did not quantify unit deformation mechanisms in compression, we observe that the NequIP model for Cu has a low flow stress, which leads to a higher tendency for healing events.

\subsection{Rare Event Sampling}
Local atomic structure is the fundamental signature of deformation mechanisms and microstructure evolution in a material.
In turn, the interatomic potential calculates these fine scale structure differences as changes in the potential energy, which makes it a powerful tool to collapse the high-dimensional response of a material into distributions of atomic energies.
Beyond the harmonic (i.e., elastic constants) and leading anharmonic (i.e., thermal expansion) comparisons of the potential energy surfaces, plastic deformation leads to new local minima at higher atomic energies that highlight model differences.
These subtle differences are what connect a model to its experimental counterpart and move predictions of general trends toward characteristic responses of specific materials.

\begin{figure}[t]
  \centering
  \includegraphics[width=0.50\linewidth]{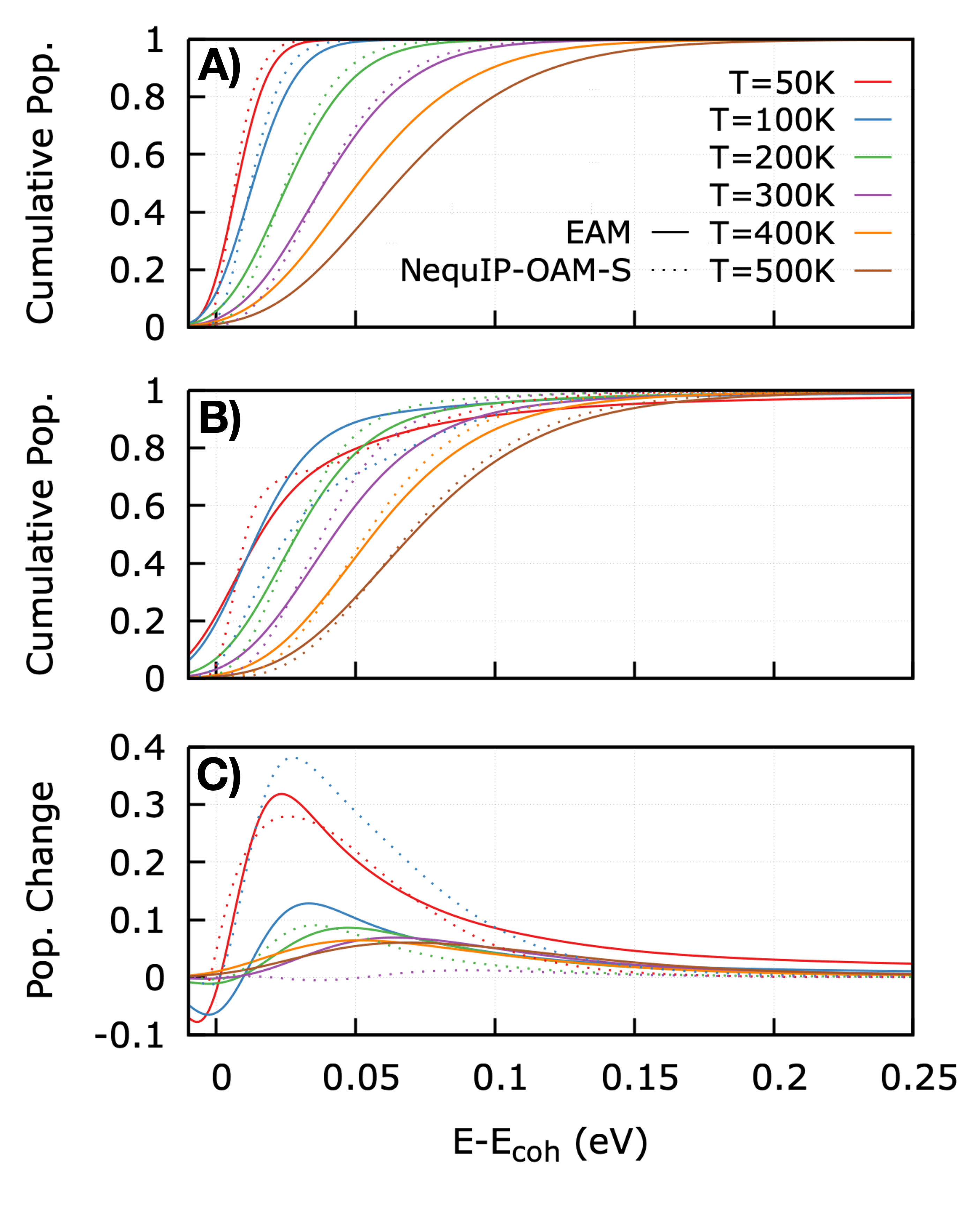}
  \caption{Cumulative distribution of normalized per-atom energies where A) the flaw does not increase in size after 10 cycles (Nominal), and B) a through crack formed, causing mechanical failure. C) Difference between the Nominal and Failure populations, averaged over repeat outcomes at the same temperature. Between potential types the largest differences are at energies above what is thermally accessible.}
  \label{fig:rare}
\end{figure}

At the end of each fatigue simulation, the potential energy distribution is gathered and offset by the cohesive energy ($E_{coh}$) of each model.
Data in Figure \ref{fig:rare} A) and B) show simulation outcomes that are classified as Nominal and Fractured, respectively, and are averaged over tensile directions and pre-stress values.
However, there are no nominal results for NequIP at $\geq400K$.
As the material deforms, the tails of this cumulative distribution clearly extend athermally to higher energy states.
This effect is more pronounced at lower temperatures, especially for T = 50~K, where both models show at least a quarter of all atoms at energies an order of magnitude higher than thermal fluctuations ($10 k_B T=0.0431$~eV).
The failure cases in Fig. \ref{fig:rare}~B) highlight differences between EAM and NequIP when $T\leq300$~K.
States accessed during deformation are energetically far above what is seen thermally, and as such we provide the difference between Nominal and Fractured cases in Fig. \ref{fig:rare}~C).
We see that NequIP predicts a population increase just above what is thermally accessible ($0.025\leq E-E_{coh}\leq0.15$ eV), whereas EAM predicts a longer high-energy tail to this distribution at all temperatures.
Qualitative comparisons between potentials based on the global outcome of the fatigue cycling are backed by the quantitative comparisons of per-atom energy distributions.
Rare event phenomena are crucial to compare interatomic potentials, as many of the predominant use cases for molecular dynamics over a higher-accuracy method like DFT are to capture transient/dynamic responses or where the simulation is used as an unbiased sampling technique.
Training sets for foundational IAP like NequIP are built from crystal structure searches\cite{pickard2011ab, jain2013commentary} and, increasingly, from rattled and high-temperature AIMD configurations\cite{barros2026open}, but they are not designed to include the defect configurations that arise during deformation, such as dislocation nucleation at a crack tip.
The purpose of these Griffith crack simulations is not to provide predictions of fracture mechanics, as this is already well studied for Cu.
Rather, the flaw geometry and thermodynamic conditions sampled here promote exploration of the simulation trajectories into local minima that are non-intuitive to construct \textit{a priori} for comparison.
Dislocation emission and interactions, which are key for understanding the strength of the material, proceed naturally from these simulations.
Taken together with the computational performance metrics and unit properties, these comparisons between cutting-edge foundational interatomic potentials and well-vetted, physically constructed models pave new ground toward widespread utilization of these model forms.
These results are also timely given the highly competitive nature of constructing foundational IAP\cite{riebesell2025framework}, which is focused on fitting metrics such as energy and force RMSE with respect to DFT.

\subsection{Stability and Extrapolation}
Before closing, we want to stress two additional complications of using foundational interatomic potentials such as NequIP: hardware memory requirements and extrapolation mitigation.
The set of fatigue simulations utilizing NequIP were run on a combination of NERSC Perlmutter and OLCF Frontier resources, each equipped with at least 64~GB per GPU (NVIDIA A100 and AMD MI250X).%
Even with this leadership computing hardware and efficient implementation in LAMMPS, we  experienced a high rate of diverging simulations due to an inability to allocate sufficient memory.
These are caused by atoms moving in and out of domains assigned to a given processor, and are not completely solved by dynamic re-balancing even at 1ps intervals.
Restart files are a solution to semi-frequent crashes, but the I/O step also adds cost to the overall throughput.
As such, falling back to a restart after a crash means wasted compute on the steps taken after the last checkpoint.
Users should perform their own stability testing to determine a suitable restart checkpoint frequency.

Lastly, a risk for any machine-learned interatomic potential is prediction beyond the bounds of the training set.
These extrapolations are not detectable from a model's predictions as the training set is not queried during use.
A committee of models approach\cite{smith2018less,liu2026mixture} is a solution to this, but each of the foundational IAP is shipped as a standalone entity.
Directly inspecting these complex functions for obvious nonphysical behavior cannot be performed exhaustively, but an example of extrapolative predictions is captured in Supplemental Figure \ref{fig:SI-EOS}.
Here a perfect crystal is expanded without relaxing atom positions to the point that graph features are constructed with fewer neighboring atoms than the ideal crystal.
Clearly, there are no training data (as is reasonable) to constrain the NN at these low-density states, and spurious unphysical fluctuations in the atomic energy are produced.
Again, care must be taken by end users to dissect the potential or construct meaningful comparisons to mitigate silent extrapolative predictions.

\section{Conclusion}
The present work demonstrates the predictive capability of recently developed foundational interatomic potentials against an EAM model of Cu by comparing both equilibrium properties and rare events brought about through fatigue cycling and deformation.
As expected, we find that equilibrium property comparisons such as the USFE and ISFE provide explanations of the observed deformation differences, but do not account for population changes in high-energy states after fatigue.
While the NequIP foundational potentials offer transferability to nearly any material class, the quantified compute cost and runtime failures are a warning to future users of these models.
Future efforts that utilize foundational potentials should construct comparisons to physically bounded models where possible, as their low relative compute cost enables valuable validation.

\section*{Acknowledgments}
{
All the authors in this work are supported by the U.S\@.~Department of Energy, Office of Science, Office of Advanced Scientific Computing Research and  Office of Basic Energy Sciences, Scientific Discovery through Advanced Computing (SciDAC) program under the MIRAGE project, award DE-SCL0000066.
This research used resources of the National Energy Research Scientific Computing Center (NERSC), a Department of Energy User Facility using NERSC award BES-ERCAP0035465 and resources at the Oak Ridge Leadership Computing Facility at the Oak Ridge National Laboratory, which is supported by the Office of Science of the U.S. Department of Energy under Contract No. DE-AC05-00OR22725.
The initial template of atomistic simulation was developed at the Center for Integrated Nanotechnologies (CINT), an Office of Science user facility operated for the U.S\@.~Department of Energy.
This article has been authored by an employee of National Technology \& Engineering Solutions of Sandia, LLC under Contract No\@.~DE-NA0003525 with the U.S\@.~Department of Energy (DOE).
The employee owns all right, title, and interest in and to the article and is solely responsible for its contents. The United States Government retains and the publisher, by accepting the article for publication, acknowledges that the United States Government retains a non-exclusive, paid-up, irrevocable, world-wide license to publish or reproduce the published form of this article or allow others to do so, for United States Government purposes.
The DOE will provide public access to these results of federally sponsored research in accordance with the DOE Public Access Plan {https://www.energy.gov/downloads/doe-public-access-plan}.

}

\bibliographystyle{unsrt}
\bibliography{references}

\begin{thebibliography}{10}

\bibitem{jumper2021highly}
John Jumper, Richard Evans, Alexander Pritzel, Tim Green, Michael Figurnov,
  Olaf Ronneberger, Kathryn Tunyasuvunakool, Russ Bates, Augustin
  {\v{Z}}{\'\i}dek, Anna Potapenko, et~al.
\newblock Highly accurate protein structure prediction with alphafold.
\newblock {\em nature}, 596(7873):583--589, 2021.

\bibitem{bertoline2023before}
Let{\'\i}cia~MF Bertoline, Ang{\'e}lica~N Lima, Jose~E Krieger, and Samantha~K
  Teixeira.
\newblock Before and after alphafold2: An overview of protein structure
  prediction.
\newblock {\em Frontiers in bioinformatics}, 3:1120370, 2023.

\bibitem{durrant2011molecular}
Jacob~D Durrant and J~Andrew McCammon.
\newblock Molecular dynamics simulations and drug discovery.
\newblock {\em BMC biology}, 9(1):71, 2011.

\bibitem{friesner2004glide}
Richard~A Friesner, Jay~L Banks, Robert~B Murphy, Thomas~A Halgren, Jasna~J
  Klicic, Daniel~T Mainz, Matthew~P Repasky, Eric~H Knoll, Mee Shelley, Jason~K
  Perry, et~al.
\newblock Glide: a new approach for rapid, accurate docking and scoring. 1.
  method and assessment of docking accuracy.
\newblock {\em Journal of medicinal chemistry}, 47(7):1739--1749, 2004.

\bibitem{laio2008metadynamics}
Alessandro Laio and Francesco~L Gervasio.
\newblock Metadynamics: a method to simulate rare events and reconstruct the
  free energy in biophysics, chemistry and material science.
\newblock {\em Reports on Progress in Physics}, 71(12):126601, 2008.

\bibitem{ceriotti2011simplifying}
Michele Ceriotti, Gareth~A Tribello, and Michele Parrinello.
\newblock Simplifying the representation of complex free-energy landscapes
  using sketch-map.
\newblock {\em Proceedings of the National Academy of Sciences},
  108(32):13023--13028, 2011.

\bibitem{kastner2011umbrella}
Johannes K{\"a}stner.
\newblock Umbrella sampling.
\newblock {\em Wiley Interdisciplinary Reviews: Computational Molecular
  Science}, 1(6):932--942, 2011.

\bibitem{zepeda2017probing}
Luis~A Zepeda-Ruiz, Alexander Stukowski, Tomas Oppelstrup, and Vasily~V
  Bulatov.
\newblock Probing the limits of metal plasticity with molecular dynamics
  simulations.
\newblock {\em Nature}, 550(7677):492--495, 2017.

\bibitem{zepeda2021atomistic}
Luis~A Zepeda-Ruiz, Alexander Stukowski, Tomas Oppelstrup, Nicolas Bertin,
  Nathan~R Barton, Rodrigo Freitas, and Vasily~V Bulatov.
\newblock Atomistic insights into metal hardening.
\newblock {\em Nature materials}, 20(3):315--320, 2021.

\bibitem{wolf2022deformation}
D~Wolf, V~Yamakov, SR~Phillpot, and AK~Mukherjee.
\newblock Deformation mechanism and inverse hall--petch behavior in
  nanocrystalline materials.
\newblock {\em International Journal of Materials Research}, 94(10):1091--1097,
  2022.

\bibitem{schiotz2003maximum}
Jakob Schi{\o}tz and Karsten~W Jacobsen.
\newblock A maximum in the strength of nanocrystalline copper.
\newblock {\em Science}, 301(5638):1357--1359, 2003.

\bibitem{schiotz1998softening}
Jakob Schi{\o}tz, Francesco~D Di~Tolla, and Karsten~W Jacobsen.
\newblock Softening of nanocrystalline metals at very small grain sizes.
\newblock {\em Nature}, 391(6667):561--563, 1998.

\bibitem{jacobs2025practical}
Ryan Jacobs, Dane Morgan, Siamak Attarian, Jun Meng, Chen Shen, Zhenghao Wu,
  Clare~Yijia Xie, Julia~H Yang, Nongnuch Artrith, Ben Blaiszik, et~al.
\newblock A practical guide to machine learning interatomic potentials--status
  and future.
\newblock {\em Current Opinion in Solid State and Materials Science},
  35:101214, 2025.

\bibitem{eyert2023machine}
Volker Eyert, Jonathan Wormald, William~A Curtin, and Erich Wimmer.
\newblock Machine-learned interatomic potentials: Recent developments and
  prospective applications.
\newblock {\em Journal of Materials Research}, 38(24):5079--5094, 2023.

\bibitem{lennard1937critical}
John~Edward Lennard-Jones and Albert~F Devonshire.
\newblock Critical phenomena in gases. i.
\newblock {\em Proceedings of the Royal Society of London. Series A,
  Mathematical and Physical Sciences}, 163(912):53--70, 1937.

\bibitem{paskin1972molecular}
Arthur Paskin and GJ~Dienes.
\newblock Molecular dynamic simulations of shock waves in a three-dimensional
  solid.
\newblock {\em Journal of Applied Physics}, 43(4):1605--1610, 1972.

\bibitem{wang2020lennard}
Xipeng Wang, Sim{\'o}n Ram{\'\i}rez-Hinestrosa, Jure Dobnikar, and Daan
  Frenkel.
\newblock The lennard-jones potential: when (not) to use it.
\newblock {\em Physical Chemistry Chemical Physics}, 22(19):10624--10633, 2020.

\bibitem{finnis1989interatomic}
Michael~W Finnis.
\newblock Interatomic forces, 1989.

\bibitem{hugginsInteratomicDistancesCrystals1933}
Maurice~L. Huggins and Joseph~E. Mayer.
\newblock Interatomic {{Distances}} in {{Crystals}} of the {{Alkali Halides}}.
\newblock {\em The Journal of Chemical Physics}, 1(9):643--646, September 1933.

\bibitem{daw1993embedded}
Murray~S Daw, Stephen~M Foiles, and Michael~I Baskes.
\newblock The embedded-atom method: a review of theory and applications.
\newblock {\em Materials Science Reports}, 9(7-8):251--310, 1993.

\bibitem{brenner2000art}
Donald~W Brenner.
\newblock The art and science of an analytic potential.
\newblock {\em physica status solidi (b)}, 217(1):23--40, 2000.

\bibitem{brenner2002second}
Donald~W Brenner, Olga~A Shenderova, Judith~A Harrison, Steven~J Stuart, Boris
  Ni, and Susan~B Sinnott.
\newblock A second-generation reactive empirical bond order (rebo) potential
  energy expression for hydrocarbons.
\newblock {\em Journal of Physics: Condensed Matter}, 14(4):783--802, 2002.

\bibitem{jelinek2012modified}
Bohumir Jelinek, Sebastien Groh, Mark~F Horstemeyer, Jeffery Houze, Seong-Gon
  Kim, Gregory~J Wagner, Amitava Moitra, and Michael~I Baskes.
\newblock Modified embedded atom method potential for al, si, mg, cu, and fe
  alloys.
\newblock {\em Physical Review B—Condensed Matter and Materials Physics},
  85(24):245102, 2012.

\bibitem{rappe1991charge}
Anthony~K Rappe and William~A Goddard~III.
\newblock Charge equilibration for molecular dynamics simulations.
\newblock {\em The Journal of Physical Chemistry}, 95(8):3358--3363, 1991.

\bibitem{van2001reaxff}
Adri~CT Van~Duin, Siddharth Dasgupta, Francois Lorant, and William~A Goddard.
\newblock Reaxff: a reactive force field for hydrocarbons.
\newblock {\em The Journal of Physical Chemistry A}, 105(41):9396--9409, 2001.

\bibitem{plimpton2012computational}
Steven~J Plimpton and Aidan~P Thompson.
\newblock Computational aspects of many-body potentials.
\newblock {\em MRS bulletin}, 37(5):513--521, 2012.

\bibitem{zuo2020performance}
Yunxing Zuo, Chi Chen, Xiangguo Li, Zhi Deng, Yiming Chen, J{\"o}rg Behler,
  G{\'a}bor Cs{\'a}nyi, Alexander~V Shapeev, Aidan~P Thompson, Mitchell~A Wood,
  et~al.
\newblock Performance and cost assessment of machine learning interatomic
  potentials.
\newblock {\em The Journal of Physical Chemistry A}, 124(4):731--745, 2020.

\bibitem{riebesell2025framework}
Janosh Riebesell, Rhys~EA Goodall, Philipp Benner, Yuan Chiang, Bowen Deng,
  Gerbrand Ceder, Mark Asta, Alpha~A Lee, Anubhav Jain, and Kristin~A Persson.
\newblock A framework to evaluate machine learning crystal stability
  predictions.
\newblock {\em Nature Machine Intelligence}, 7(6):836--847, 2025.

\bibitem{jiaPushingLimitMolecular2020}
Weile Jia, Han Wang, Mohan Chen, Denghui Lu, Lin Lin, Roberto Car, Weinan E,
  and Linfeng Zhang.
\newblock Pushing the {{Limit}} of {{Molecular Dynamics}} with {{Ab Initio
  Accuracy}} to 100 {{Million Atoms}} with {{Machine Learning}}.
\newblock In {\em {{SC20}}: {{International Conference}} for {{High Performance
  Computing}}, {{Networking}}, {{Storage}} and {{Analysis}}}, pages 1--14,
  Atlanta, GA, USA, November 2020. IEEE.

\bibitem{nguyen2021billion}
Kien Nguyen-Cong, Jonathan~T Willman, Stan~G Moore, Anatoly~B Belonoshko,
  Rahulkumar Gayatri, Evan Weinberg, Mitchell~A Wood, Aidan~P Thompson, and
  Ivan~I Oleynik.
\newblock Billion atom molecular dynamics simulations of carbon at extreme
  conditions and experimental time and length scales.
\newblock In {\em Proceedings of the International Conference for High
  Performance Computing, Networking, Storage and Analysis}, pages 1--12, 2021.

\bibitem{kozinsky2023scaling}
Boris Kozinsky, Albert Musaelian, Anders Johansson, and Simon Batzner.
\newblock Scaling the leading accuracy of deep equivariant models to
  biomolecular simulations of realistic size.
\newblock In {\em Proceedings of the International conference for high
  performance computing, networking, storage and analysis}, pages 1--12, 2023.

\bibitem{smith2019approaching}
Justin~S Smith, Benjamin~T Nebgen, Roman Zubatyuk, Nicholas Lubbers, Christian
  Devereux, Kipton Barros, Sergei Tretiak, Olexandr Isayev, and Adrian~E
  Roitberg.
\newblock Approaching coupled cluster accuracy with a general-purpose neural
  network potential through transfer learning.
\newblock {\em Nature communications}, 10(1):2903, 2019.

\bibitem{mattsson2005designing}
Ann~E Mattsson, Peter~A Schultz, Michael~P Desjarlais, Thomas~R Mattsson, and
  Kevin Leung.
\newblock Designing meaningful density functional theory calculations in
  materials science—a primer.
\newblock {\em Modelling and Simulation in Materials Science and Engineering},
  13(1):R1--R31, 2005.

\bibitem{jaspersonCrossscaleCovarianceMaterial2025}
Benjamin~A. Jasperson, Ilia Nikiforov, Amit Samanta, Fei Zhou, Ellad~B. Tadmor,
  Vincenzo Lordi, and Vasily~V. Bulatov.
\newblock Cross-scale covariance for material property prediction.
\newblock {\em npj Computational Materials}, 11(1):1, January 2025.

\bibitem{jaspersonFundamentalMicroscopicProperties2025}
Benjamin~A. Jasperson, Ilia Nikiforov, Amit Samanta, Brandon Runnels, Harley~T.
  Johnson, and Ellad~B. Tadmor.
\newblock Fundamental microscopic properties as predictors of large-scale
  quantities of interest: {{Validation}} through grain boundary energy trends.
\newblock {\em Acta Materialia}, 286:120722, March 2025.

\bibitem{dingreville2017primer}
R{\'e}mi Dingreville, Doruk Aksoy, and Douglas~E Spearot.
\newblock A primer on selecting grain boundary sets for comparison of
  interfacial fracture properties in molecular dynamics simulations.
\newblock {\em Scientific Reports}, 7(1):8332, 2017.

\bibitem{fu2022forces}
Xiang Fu, Zhenghao Wu, Wujie Wang, Tian Xie, Sinan Keten, Rafael
  Gomez-Bombarelli, and Tommi Jaakkola.
\newblock Forces are not enough: Benchmark and critical evaluation for machine
  learning force fields with molecular simulations.
\newblock {\em arXiv preprint arXiv:2210.07237}, 2022.

\bibitem{sikorski2023machine}
EL~Sikorski, MA~Cusentino, MJ~McCarthy, J~Tranchida, MA~Wood, and AP~Thompson.
\newblock Machine learned interatomic potential for dispersion strengthened
  plasma facing components.
\newblock {\em The Journal of Chemical Physics}, 158(11), 2023.

\bibitem{perez2024recent}
Danny Perez and Tony Leli{\`e}vre.
\newblock Recent advances in accelerated molecular dynamics methods: Theory and
  applications.
\newblock {\em Comprehensive Computational Chemistry}, 3:360--383, 2024.

\bibitem{barr2023autonomous}
Christopher~M Barr, Ta~Duong, Daniel~C Bufford, Zachary Milne, Abhilash
  Molkeri, Nathan~M Heckman, David~P Adams, Ankit Srivastava, Khalid Hattar,
  Michael~J Demkowicz, et~al.
\newblock Autonomous healing of fatigue cracks via cold welding.
\newblock {\em Nature}, 620(7974):552--556, 2023.

\bibitem{boyce2026materials}
Brad~L Boyce, Mitchell~A Wood, Krishna Garikipati, Andreas~E Robertson, Jeffrey
  Larson, Ishan Srivastava, Bert Debusschere, Saaketh Desai, Prasad Iyer,
  Pieterjan Robbe, Mathew Cherukara, Todd Munson, Ming Du, Trupti Mohanty,
  David~J. Gardner, Laurent Capolungo, Benjamin~A. Jasperson, Jingye Tan, and
  Remi Dingreville.
\newblock Materials behavior as mechanism ensembles: A probabilistic framework
  for emergent behaviors.
\newblock {\em arXiv preprint arXiv:2607.27163}, 2026.

\bibitem{mishin2001structural}
Yuri Mishin, Michael~J Mehl, Dimitrios~A Papaconstantopoulos, Arthur~F Voter,
  and Joel~D Kress.
\newblock Structural stability and lattice defects in copper: Ab initio,
  tight-binding, and embedded-atom calculations.
\newblock {\em Physical Review B}, 63(22):224106, 2001.

\bibitem{pei2015brittle}
Linqing Pei, Cheng Lu, Xing Zhao, Liang Zhang, Kuiyu Cheng, Guillaume Michal,
  and Kiet Tieu.
\newblock Brittle versus ductile behaviour of nanotwinned copper: a molecular
  dynamics study.
\newblock {\em Acta Materialia}, 89:1--13, 2015.

\bibitem{ogata2002ideal}
Shigenobu Ogata, Ju~Li, and Sidney Yip.
\newblock Ideal pure shear strength of aluminum and copper.
\newblock {\em Science}, 298(5594):807--811, 2002.

\bibitem{zeng2015fracture}
Zhi Zeng, Xiaoyan Li, Lei Lu, and Ting Zhu.
\newblock Fracture in a thin film of nanotwinned copper.
\newblock {\em Acta Materialia}, 98:313--317, 2015.

\bibitem{cao2008grain}
Ajing Cao, Yueguang Wei, and En~Ma.
\newblock Grain boundary effects on plastic deformation and fracture mechanisms
  in cu nanowires: Molecular dynamics simulations.
\newblock {\em Physical Review B—Condensed Matter and Materials Physics},
  77(19):195429, 2008.

\bibitem{batzner20223}
Simon Batzner, Albert Musaelian, Lixin Sun, Mario Geiger, Jonathan~P Mailoa,
  Mordechai Kornbluth, Nicola Molinari, Tess~E Smidt, and Boris Kozinsky.
\newblock E (3)-equivariant graph neural networks for data-efficient and
  accurate interatomic potentials.
\newblock {\em Nature communications}, 13(1):2453, 2022.

\bibitem{kavanagh2026fast}
Se{\'a}n~R Kavanagh, Chuin~Wei Tan, Menghang Wang, Marc~L Descoteaux, Gabriel
  de~Miranda Nascimento, Ulrik Unneberg, Laura Zichi, Francesco Libbi, Norma
  Rivano, Austin Glover, et~al.
\newblock Fast and accurate foundation models for equivariant machine-learned
  interatomic potentials.
\newblock {\em arXiv preprint arXiv:2607.28461}, 2026.

\bibitem{barros2026open}
Luis Barros-Luque, Muhammed Shuaibi, Xiang Fu, Brandon~M Wood, Misko Dzamba,
  Meng Gao, Ammar Rizvi, Matt Uyttendaele, C~Lawrence Zitnick, and Zachary~W
  Ulissi.
\newblock The open materials 2024 (omat24) inorganic materials dataset and
  models.
\newblock {\em Nature Computational Science}, pages 1--11, 2026.

\bibitem{schmidt2023machine}
Jonathan Schmidt, Noah Hoffmann, Hai-Chen Wang, Pedro Borlido, Pedro~JMA
  Carri{\c{c}}o, Tiago~FT Cerqueira, Silvana Botti, and Miguel~AL Marques.
\newblock Machine-learning-assisted determination of the global
  zero-temperature phase diagram of materials.
\newblock {\em Advanced Materials}, 35(22):2210788, 2023.

\bibitem{wang2023symmetry}
Hai-Chen Wang, Jonathan Schmidt, Miguel~AL Marques, Ludger Wirtz, and Aldo~H
  Romero.
\newblock Symmetry-based computational search for novel binary and ternary 2d
  materials.
\newblock {\em 2D Materials}, 10(3):035007, 2023.

\bibitem{dengCHGNetPretrainedUniversal2023}
Bowen Deng, Peichen Zhong, KyuJung Jun, Janosh Riebesell, Kevin Han,
  Christopher~J. Bartel, and Gerbrand Ceder.
\newblock {{CHGNet}} as a pretrained universal neural network potential for
  charge-informed atomistic modelling.
\newblock {\em Nature Machine Intelligence}, 5(9):1031--1041, September 2023.

\bibitem{ziegler1985stopping}
James~F Ziegler and Jochen~P Biersack.
\newblock The stopping and range of ions in matter.
\newblock In {\em Treatise on heavy-ion science: volume 6: astrophysics,
  chemistry, and condensed matter}, pages 93--129. Springer, 1985.

\bibitem{thompson2022lammps}
Aidan~P Thompson, H~Metin Aktulga, Richard Berger, Dan~S Bolintineanu,
  W~Michael Brown, Paul~S Crozier, Pieter~J In't~Veld, Axel Kohlmeyer, Stan~G
  Moore, Trung~Dac Nguyen, et~al.
\newblock Lammps-a flexible simulation tool for particle-based materials
  modeling at the atomic, meso, and continuum scales.
\newblock {\em Computer physics communications}, 271:108171, 2022.

\bibitem{griffithVIPhenomenaRupture1921}
Alan~Arnold Griffith.
\newblock {{VI}}. {{The}} phenomena of rupture and flow in solids.
\newblock {\em Philosophical Transactions of the Royal Society of London.
  Series A, Containing Papers of a Mathematical or Physical Character},
  221(582-593):163--198, January 1921.

\bibitem{wilson2019continuum}
Mark~A Wilson, Scott~J Grutzik, and Michael Chandross.
\newblock Continuum stress intensity factors from atomistic fracture
  simulations.
\newblock {\em Computer Methods in Applied Mechanics and Engineering},
  354:732--749, 2019.

\bibitem{stukowski2010visualization}
Alexander Stukowski.
\newblock Visualization and analysis of atomistic simulation data with
  ovito--the open visualization tool.
\newblock {\em Modelling and simulation in materials science and engineering},
  18(1):015012, 2010.

\bibitem{rice1992dislocation}
James~R Rice.
\newblock Dislocation nucleation from a crack tip: an analysis based on the
  peierls concept.
\newblock {\em Journal of the Mechanics and Physics of Solids}, 40(2):239--271,
  1992.

\bibitem{li2026universal}
Tiancheng Li, Jianming Xue, Linfeng Zhang, Duo Zhang, and Han Wang.
\newblock Universal machine-learning molecular dynamics at the speed of
  empirical potentials.
\newblock {\em arXiv preprint arXiv:2608.19041}, 2026.

\bibitem{frenkel1997understanding}
Daan Frenkel, Berend Smit, and Mark~A Ratner.
\newblock Understanding molecular simulation: from algorithms to applications,
  1997.

\bibitem{Jona2001}
F.~Jona and P.~M. Marcus.
\newblock Structural properties of copper.
\newblock {\em Physical Review B}, 63(9):094113, February 2001.

\bibitem{Pike2019}
Nicholas~A. Pike and Ole~M. L{\o}vvik.
\newblock Calculation of the anisotropic coefficients of thermal expansion: A
  first-principles approach.
\newblock {\em Computational Materials Science}, 167:257--263, 2019.

\bibitem{White1973}
G.~K. White.
\newblock Thermal expansion of reference materials: copper, silica and silicon.
\newblock {\em Journal of Physics D: Applied Physics}, 6(17):2070--2078,
  November 1973.

\bibitem{Zhang2019}
Yong Zhang, Jinming Guo, Jianghua Chen, Cuilan Wu, Karl~S. Kormout, Pradipta
  Ghosh, and Zaoli Zhang.
\newblock On the stacking fault energy related deformation mechanism of
  nanocrystalline {Cu} and {Cu} alloys: A first-principles and {TEM} study.
\newblock {\em Journal of Alloys and Compounds}, 776:807--818, 2019.

\bibitem{Han2019}
Yong Han, King~C. Lai, Ann Lii-Rosales, Michael~C. Tringides, James~W. Evans,
  and Patricia~A. Thiel.
\newblock Surface energies, adhesion energies, and exfoliation energies
  relevant to copper-graphene and copper-graphite systems.
\newblock {\em Surface Science}, 685:48--58, 2019.

\bibitem{Sivak2007}
A.~B. Sivak, V.~M. Chernov, N.~A. Dubasova, and V.~A. Romanov.
\newblock Anisotropy migration of self-point defects in dislocation stress
  fields in {BCC} {Fe} and {FCC} {Cu}.
\newblock {\em Journal of Nuclear Materials}, 367--370:316--321, 2007.

\bibitem{broughton1986thermodynamic}
JQ~Broughton and GH~Gilmer.
\newblock Thermodynamic criteria for grain-boundary melting: a
  molecular-dynamics study.
\newblock {\em Physical review letters}, 56(25):2692, 1986.

\bibitem{gerward1985bulk}
L~Gerward.
\newblock The bulk modulus and its pressure derivative for 18 metals.
\newblock {\em Journal of Physics and Chemistry of Solids}, 46(8):925--927,
  1985.

\bibitem{wu2010ab}
Xiao-Zhi Wu, Rui Wang, Shao-Feng Wang, and Qun-Yi Wei.
\newblock Ab initio calculations of generalized-stacking-fault energy surfaces
  and surface energies for fcc metals.
\newblock {\em Applied Surface Science}, 256(21):6345--6349, 2010.

\bibitem{pickard2011ab}
Chris~J Pickard and RJ~Needs.
\newblock Ab initio random structure searching.
\newblock {\em Journal of Physics: Condensed Matter}, 23(5):053201, 2011.

\bibitem{jain2013commentary}
Anubhav Jain, Shyue~Ping Ong, Geoffroy Hautier, Wei Chen, William~Davidson
  Richards, Stephen Dacek, Shreyas Cholia, Dan Gunter, David Skinner, Gerbrand
  Ceder, et~al.
\newblock Commentary: The materials project: A materials genome approach to
  accelerating materials innovation.
\newblock {\em APL materials}, 1(1), 2013.

\bibitem{smith2018less}
Justin~S Smith, Ben Nebgen, Nicholas Lubbers, Olexandr Isayev, and Adrian~E
  Roitberg.
\newblock Less is more: Sampling chemical space with active learning.
\newblock {\em The Journal of chemical physics}, 148(24), 2018.

\bibitem{liu2026mixture}
Yuzhi Liu, Duo Zhang, Anyang Peng, Weinan E, Linfeng Zhang, and Han Wang.
\newblock Mixture of experts architectures for machine learning interatomic
  potentials.
\newblock {\em npj Artificial Intelligence}, 2026.

\end{thebibliography}

\pagebreak
\renewcommand\thefigure{\arabic{figure}}\setcounter{figure}{0}
\renewcommand\thetable{\arabic{table}}\setcounter{table}{0}
\setcounter{figure}{0}
\setcounter{section}{0}
\renewcommand{\figurename}{Supplemental Figure}
\renewcommand{\tablename}{Supplemental Table}

\section{Supplemental Information}

Additional detail on each of the 72 Griffith's Crack fatigue simulations that were run for EAM and NequIP-OAM-S interatomic potential types is give in Supplemental Table \ref{tab:runid}.
The 'Run ID' column is used as the color scale in Figure \ref{fig:sav} to differentiate the thermodynamic conditions for this simulations.
\begin{table}[H]
  \centering
  \caption{Enumeration of all simulation conditions for fatigue tests and assignments of unique IDs used in Figures \ref{fig:sav} and \ref{fig:rare}. Outcomes are designated as mechanical failure (F) when a through crack has formed, nominal (N) when the flaw shows minimal or no growth as measured by surface area to volume ratio, and healed (H) if the flaw volume goes to zero during fatigue cycling.}
  \label{tab:runid}
  \small
  \begin{tabular}{lcccccc}
    \toprule
    $\sigma_x$ & $\sigma_y$ & Tensile & &  & EAM & NequIP\\
    (GPa) & (GPa) & Direction & Temperature (K) & Run ID & Outcome & Outcome\\
    \midrule
    $5.0$ & $1\cdot10^{-4}$ & [100] & 50,100,200,300,400,500 & 1,2,3,4,5,6 & F,F,F,F,F,F & F,F,F,F,F,F\\
    $5.0$ & $1\cdot10^{-4}$ & [110] & 50,100,200,300,400,500 & 7,8,9,10,11,12 & N,F,F,F,F,F & F,F,F,F,F,F\\
    $5.0$ & $1\cdot10^{-4}$ & [111] & 50,100,200,300,400,500 & 13,14,15,16,17,18 & F,F,F,F,F,F & F,F,F,F,F,F\\
    \hline
    $4.0$ & $1\cdot10^{-4}$ & [100] & 50,100,200,300,400,500 & 19,20,21,22,23,24 & N,N,N,N,F,F & H,H,H,H,H,H\\
    $4.0$ & $1\cdot10^{-4}$ & [110] & 50,100,200,300,400,500 & 25,26,27,28,29,30 & N,N,F,F,F,F & N,N,F,F,H,H\\
    $4.0$ & $1\cdot10^{-4}$ & [111] & 50,100,200,300,400,500 & 31,32,33,34,35,36 & F,F,F,F,F,F & F,H,F,F,F,F\\
    \hline
    $3.0$ & $1\cdot10^{-4}$ & [100] & 50,100,200,300,400,500 & 37,38,39,40,41,42 & N,N,N,N,N,N & H,H,H,H,H,H\\
    $3.0$ & $1\cdot10^{-4}$ & [110] & 50,100,200,300,400,500 & 43,44,45,46,47,48 & N,N,N,N,F,N & N,N,N,N,H,H\\
    $3.0$ & $1\cdot10^{-4}$ & [111] & 50,100,200,300,400,500 & 49,50,51,52,53,54 & N,N,N,N,F,F & N,N,N,F,F,F\\
    \hline
    $2.0$ & $1\cdot10^{-4}$ & [100] & 50,100,200,300,400,500 & 55,56,57,58,59,60 & H,H,H,H,H,H & H,H,H,H,H,H\\
    $2.0$ & $1\cdot10^{-4}$ & [110] & 50,100,200,300,400,500 & 61,62,63,64,65,66 & N,N,N,N,N,N & H,H,H,H,H,H\\
    $2.0$ & $1\cdot10^{-4}$ & [111] & 50,100,200,300,400,500 & 67,68,69,70,71,72 & N,N,N,N,N,N & N,N,N,N,H,H\\
\bottomrule
  \end{tabular}
\end{table}

In addition to the computational efficiency data given in Figure
\ref{fig:perf} the data in Supplemental Table \ref{tab:scaling-sizes} provides additional details of system sizes.
The largest weak-scaling system is dependent on the memory per node and the storage requirements of the foundational model, hence the reduced atom count relative to EAM which is an extremely lightweight model in this regard.
\begin{table}[H]
  \centering
  \caption{Exact system sizes used in the computational efficiency benchmarks (Figure~\ref{fig:perf}). Both are cubic FCC Cu cells. In strong scaling the total atom count is fixed and run on 1 to 512 nodes; in weak scaling the same count is held fixed per node. }
  \label{tab:scaling-sizes}
  \small
  \begin{tabular}{lccc}
    \toprule
    Label & FCC unit cells & Atoms & Largest weak-scaling system (512 | 256 nodes) \\
    \midrule
    0.1M & $32^3$ & 131,072 & 67,108,864 \\
    1M & $64^3$ & 1,048,576 &  EAM: 536,870,912 | NequIP-OAM-S: 268,435,456 \\
    \bottomrule
  \end{tabular}
\end{table}

\begin{figure}[t]
  \centering
  \includegraphics[width=0.50\linewidth]{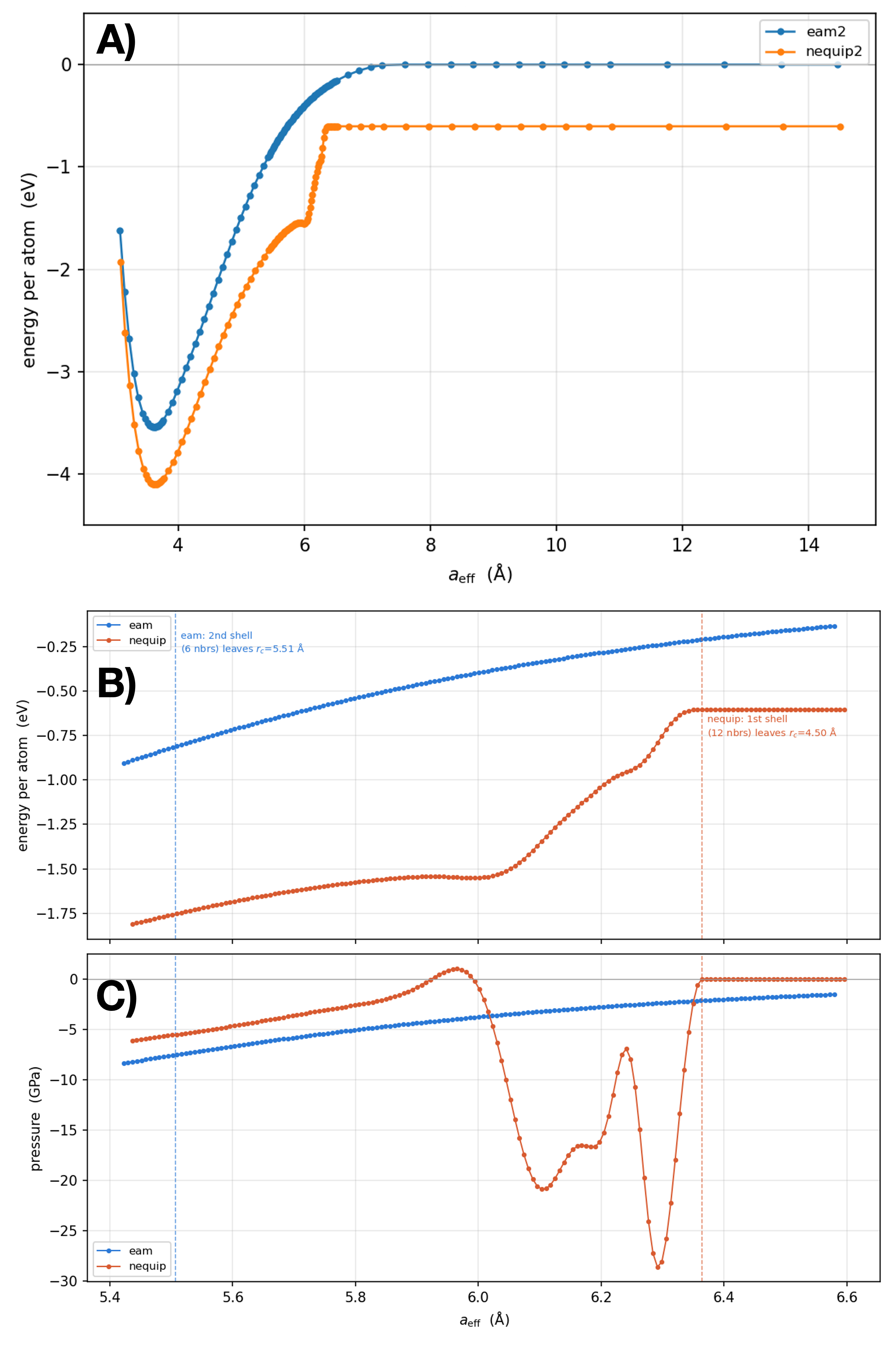}
  \caption{Inspection of the potential energy surface as FCC Cu is volumetrically expanded and contracted. A) Expansion beyond the radial cutoff of either potential results in a constant of energy predicted for the isolated atom, this transition is smooth for EAM and rapidly varying for NequIP, see panel B) for closer inspection. For NequIP this results in large pressures of few-body interactions and we deem this an extrapolation of the potential. C) The pressure, $r_{ij}~\partial/\partial r_{ij}~U(r_{ij})$, where $r_{ij}$ is the interatomic distance. \label{fig:SI-EOS}}

\end{figure}

By compressing and expanding the unit cell of a given material one can inspect a trace of the high dimensional potential energy surface, and is colloquially called the cold-curve equation of state (EOS).
Supplemental Figure \ref{fig:SI-EOS} compares the EAM and NequIP EOS where the focus is on the expanded crystal where neighbor lists of particles contain fewer than the equilibrium crystal.
The data is given as the energy per atom against the lattice constant of this modified crystal.
Predictably the EAM model shows smooth and continuous progress toward the energy of an isolated atom, which is constant after a given $a_{eff}\simeq8\AA$
By contrast the under-coordinated atom environments produce high-frequency features in the potential energy for NequIP, which are clearly extrapolations from what is contained in the training set.
For completeness the pressure, $r_{ij}~\partial/\partial r_{ij}~U(r_{ij})$, is shown in Supp. Fig. \ref{fig:SI-EOS} C) where the extrapolative predictions of NequIP show up clearly.

\end{document}